\documentstyle[aps,pre,epsf]{revtex}
\newcommand{\bm}[1]{\mbox{\boldmath$#1$}}

\begin{document}

\twocolumn[\hsize\textwidth\columnwidth\hsize\csname@twocolumnfalse\endcsname

\title{Slow Switching in Globally Coupled Oscillators:\\
Robustness and Occurrence through Delayed Coupling} 
\author{Hiroshi Kori\cite{byline} and Yoshiki Kuramoto}

\address{Department of Physics, Graduate School of Sciences, Kyoto University,
Kyoto 606-8502, Japan}

\maketitle

\widetext

\begin{abstract}
 The phenomenon of slow switching in populations of globally coupled
 oscillators is discussed.  This characteristic collective dynamics,
 which was first discovered in a particular class of the phase oscillator
 model, is a result of the formation of a heteroclinic loop connecting a
 pair of clustered states of the population. We argue that the same
 behavior can arise in a wider class of oscillator models with the
 amplitude degree of freedom. We also argue how such heteroclinic loops
 arise inevitably and persist robustly in a homogeneous population of
 globally coupled oscillators. Although the heteroclinic loop might seem
 to arise only exceptionally, we find that it appears rather easily by
 introducing the time-delay in the population which would otherwise
 exhibit perfect phase synchrony. We argue that the appearance of the
 heteroclinic loop induced by the delayed coupling is then characterized
 by transcritical and saddle-node bifurcations. Slow switching arises
 when the system with a heteroclinic loop is weakly perturbed. This will
 be demonstrated with a vector model by applying weak noises. Other
 types of weak symmetry-breaking perturbations can also cause slow
 switching.
\end{abstract}


]

\narrowtext

\section{Introduction} \label{sec:1}

\begin{figure}
\centerline{ \epsfxsize=9cm\epsfbox{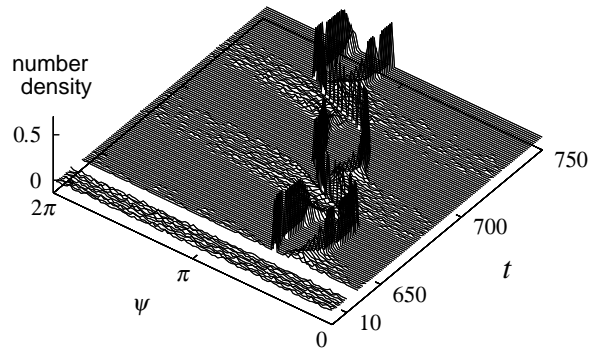}}
\caption{Slow switching exhibited by the model in [10]. The figure displays
 the time-evolution of the number density of the oscillators as a
 function of the phase. The whole population, which was initially almost
 uniform, splits into two subpopulations each almost converging to a
 point cluster. After some time, however, this seeming convergence turns
 out unstable, and is followed by the period of their scattering, but
 again this is followed by the period of the convergence, and so forth.
 This form of alternation between the two characteristic periods of the
 convergence and dispersion of the clusters is called slow switching.
 The phase-advanced/retarded cluster at the end of one cycle becomes
 phase-retarded/advanced at the end of the next cycle. } 
 \label{fig:slowswitch}
\end{figure}

Coupled limit-cycle oscillators appear in various contexts in physics
\cite{watanabe94,wiesenfeld96,hansel93-lett,okuda}, chemistry
\cite{kuramoto84,rubin94} and biology \cite{winfree67,buck88,delaney94}. 
Various types of collective behavior which arise when they form large
assemblies have been studied extensively over the last few decades. Among
the possible types of behavior, we will particularly be concerned with
{\em clustering and slow switching} which was first studied by Hansel et
al.\cite{hansel93} in a homogeneous population of globally coupled phase
oscillators.  Assuming a simple form for the coupling function, they
showed numerically that after a long transient the system approaches a
{\em two-cluster state}, i.e. the whole population splits into rigidly
rotating two subpopulations each in perfect phase synchrony. However,
the stability analysis of this two-cluster state revealed that it is
linearly unstable, corresponding to a saddle point if seen in a co-rotating
frame of reference. The seeming contradiction here was interpreted in
terms of the formation of a heteroclinic loop connecting this
two-cluster state and another two-cluster state which is obtained simply
by a constant phase shift of the former. In fact, when this heteroclinic
loop is attracting, the trajectory comes to stay longer and longer near
these saddle points, so that the numerical round-off error finally
forces the system to stay at one of the saddles forever.  This
interpretation was supported by the fact that when small external noise
is included the system is no longer fixed at a saddle point but starts
to repeat slow switchings between the pair of saddles (see Fig.\
\ref{fig:slowswitch}). Although these findings are so important,
explanation of the reason is still needed why the heteroclinic loop
arises inevitably and persists robustly against our common belief in its
structural instability.

In the next two sections, we restrict our consideration to the phase
model.  In Sec. \ref{sec:2}, we discuss in some detail the stability
condition of the two-cluster state. Existing
\noindent
numerical results obtained by a particular model suggest the appearance
of heteroclinic loops.  Thus, in Sec. \ref{sec:3}, we argue the
mechanism by which heteroclinic loops are necessarily formed.
Specifically, a sufficient condition will be given for the existence of
a heteroclinic loop, and how this condition is satisfied in the phase
model will be discussed. In Sec. \ref{sec:4}, we introduce a specific
vector oscillator model for globally coupled oscillators and exhibit
numerically that heteroclinic loops are formed in our vector model. We
show there that the phase-coupling function, derived numerically from
the vector model by the method of the phase reduction, satisfies the
above-mentioned condition leading to the formation of heteroclinic
loops.  Our result gives the first result showing the heteroclinic loop
in the vector model of coupled oscillators.  In Sec. \ref{sec:5}, we
generalize the argument in Sec. \ref{sec:3} to the vector model.

The formation of heteroclinic loops in globally coupled oscillators may
seem to be a pathological phenomenon which can happen only
exceptionally.  However, the time-delay in coupling can easily cause a
bifurcation from perfect synchrony to the formation of a heteroclinic
loop, and this will be discussed in Sec. \ref{sec:6}. The method of the 
phase reduction provides a clear understanding of why this is actually
possible.  Slow switching becomes persistent when the oscillators are
subject to weak external noise, which will be discussed in
Sec. \ref{sec:7} by using a vector model.  We will also show there that
the same phenomenon can also be caused by other types of randomness.

\section{Heteroclinic loop in the phase model} \label{sec:2}
Populations of weakly coupled limit cycle oscillators can be described 
by the phase model\cite{kuramoto84}.  Suppose that the oscillators
are identical, each interacting with all the others with equal strength.
Then the corresponding phase model is expressed as 
\begin{equation}
 \frac{d}{dt}\psi_{i}(t) = \omega + \frac{K}{N}
	\sum^{N}_{j=1}\Gamma [ \psi_i(t)-\psi_j(t) ] ,
\label{pm}
\end{equation}
where $\psi_i(t)\ (0 \leq \psi_i < 2\pi)$ is the phase of the $i$-th
oscillator $(i=1,\cdots,N)$, $\omega$ and $K$ are positive constants,
and $\Gamma(x)$ is the coupling function with $2\pi$-periodicity.
\begin{figure}
\centerline{\epsfxsize=9cm\epsfbox{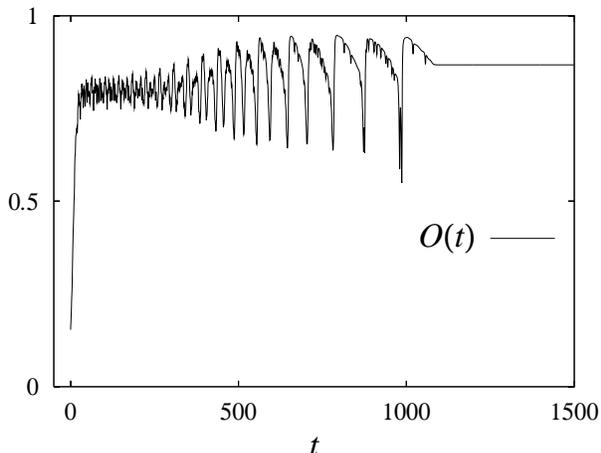}}
  \caption{ Long transient of the order parameter after which the whole
  population converges to a two-cluster state.}  \label{fig:order-hansel}
\end{figure}

\begin{figure}
\epsfxsize=7.5cm\epsfbox{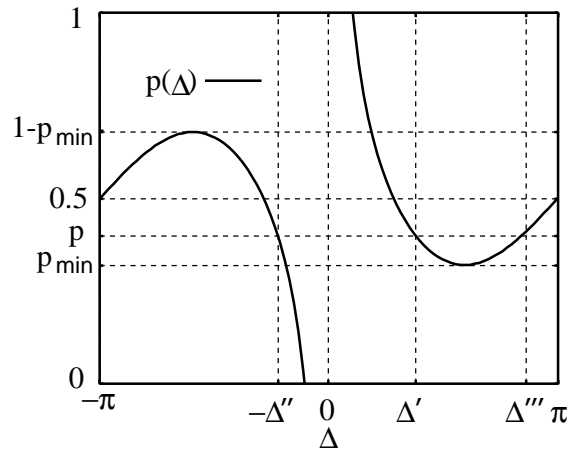}
  \caption{ Condition for the existence of two-cluster states. $\Delta$
  takes three values for a given $p$ within the range $p_{\rm min} < p <
  1-p_{\rm min}$. } \label{fig:p-delta-hansel}
\end{figure}

Hansel et al.\ \cite{hansel93} analyzed the case of a particular
form of the coupling function 
\begin{equation}
 \Gamma(x)=-\sin(x+1.25)+0.25\sin(2x).
 \label{gamma-hansel}
\end{equation}
They showed by numerical simulations that the oscillators with random
initial distribution are assembled to form two subgroups each in perfect
phase synchrony but with a constant mutual phase difference. The
collective behavior of the system can conveniently be described in terms
of the order parameter defined by
\begin{equation}
 O(t)=\frac{1}{N} \left| \sum_{j=1}^{N} \exp[i \psi_j] \right|.
\end{equation}
Its value is $1$ for perfect synchrony and $0$ for perfect incoherence.
A time trace of the order parameter for the
above model is displayed in Fig.\ \ref{fig:order-hansel}.  The
oscillators belonging to the respective groups are identical in phase,
and this pair of point clusters rotate rigidly at a constant angular
frequency.  The mutual phase difference is denoted by $\Delta$. We call
hereafter the phase-advanced and retarded clusters the A-cluster and
B-cluster, respectively.  Let the fraction of the oscillators belonging
to the A-cluster be $p$. Such a two-cluster state may thus be specified
by $(p,\Delta)$, where $\Delta$ is within the region $-\pi< \Delta \leq
\pi$ by convention. This set of values may generally differ for
different initial conditions.

Existence and stability of the two-cluster states can be analyzed as
follows. Consider a two-cluster state with phases $\psi_{\rm A}$ and
$\psi_{\rm B}$.  Eq. (\ref{pm}) then becomes
\begin{equation} \label{pm-A}
 \frac{d}{dt}\psi_{\rm A}(t) = \omega + K\{ p\Gamma(0)
	+(1-p)\Gamma(x) \}, 
\end{equation}
\begin{equation} \label{pm-B}
 \frac{d}{dt}\psi_{\rm B}(t) = \omega + K\{ (1-p)\Gamma(0)
	+p\Gamma(-x) \},
\end{equation}
where $x$ denotes the phase difference, i.e. $x \equiv \psi_{\rm
A}-\psi_{\rm B}$. Subtracting Eq. (\ref{pm-B}) from Eq. (\ref{pm-A}), we
obtain
\begin{figure}
\epsfxsize=8.5cm\epsfbox{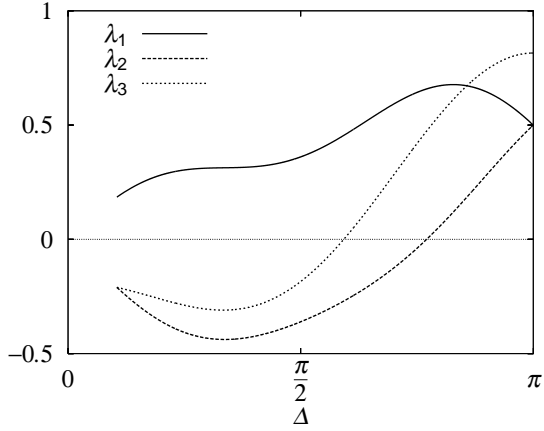}
  \caption{ Eigenvalues about two-cluster states as a function of
  $\Delta$. All two-cluster states are unstable here. Note that the
  eigenvalues $\lambda_2$ and $\lambda_3$ are negative for the states
  $(p,\Delta')$ and $(1-p,\Delta'')$. } \label{fig:eigen-hansel}
\end{figure}
\noindent
\begin{equation} \label{pm-x}
 \frac{d}{dt}x(t) = K\{ (2p-1)\Gamma(0)+(1-p)\Gamma(x)-p\Gamma(-x) \}.
\end{equation}
Since $x$ is constant in the two-cluster state, we have
 \begin{equation} \label{p-delta}
 p(\Delta)=\frac{\Gamma(0)-\Gamma(\Delta)}
  {2\Gamma(0)-\Gamma(\Delta)-\Gamma(-\Delta)}.
\end{equation}
$(p,\Delta)$ exists with $p$ satisfying $0<p<1$. Substituting
Eq. (\ref{gamma-hansel}) into Eq. (\ref{p-delta}), we obtain
the
condition for the existence of $(p,\Delta)$, and this is displayed
graphically in Fig.\ \ref{fig:p-delta-hansel}. $\Delta$ takes three
values for a given $p$ within the range $p_{\rm{min}} < p <
1-p_{\rm{min}}$, where $p_{\rm{min}}$ is defined by the minimum value of
$p$ in the range $0<\Delta<\pi$. These three states are denoted by $(p,
\Delta')$, $(p, -\Delta'')=(1-p, \Delta'')$ and $(p,\Delta''')$, where
$\Delta'$ and $\Delta''$ are understood to be positive and $|\Delta'''|$
to be larger than $\Delta'$ and $\Delta''$.

The eigenvalues of the stability matrix
are given by
\begin{equation}\label{l0}
  \lambda_{0} = 0, 
\end{equation}
\begin{equation}\label{l1}
  \lambda_{1} = K\{p\Gamma'(0)+(1-p)\Gamma'(\Delta)\}, 
\end{equation}
\begin{equation}\label{l2}
  \lambda_{2} = K\{(1-p)\Gamma'(0)+p\Gamma'(-\Delta)\},  
\end{equation}
\begin{equation} \label{l3}
  \lambda_{3} = K\{(1-p)\Gamma'(\Delta)+p\Gamma'(-\Delta)\},
\end{equation}
with multiplicity $1$,$Np-1$,$N(1-p)-1$ and $1$ respectively. 
$\Gamma'(x)$ is defined as $\frac{d}{dx}\Gamma(x)$.  $\lambda_{0}$, which
vanishes identically, always exists due to the invariance of Eq.\ (4)
with respect to a constant shift of $\psi_A$ and $\psi_B$.
$\lambda_{1}$ and $\lambda_{2}$ are associated with the fluctuations of
the individual oscillators belonging to the A-cluster and B-cluster,
respectively.  $\lambda_{3}$ corresponds to the fluctuation in
$\Delta$. Figure \ref{fig:eigen-hansel} displays the eigenvalues versus
$\Delta$ obtained using (\ref{gamma-hansel}) with $K=1$, which shows that
all two-cluster states are unstable. It is important to note that
$(p,\Delta')$ and $(1-p,\Delta'')$ are saddles which have negative
eigenvalues of $\lambda_2$ and $\lambda_3$. $(p,\Delta''')$, however,
has a positive $\lambda_3$, which can be verified by the property
$\lambda_3 \propto \frac{d}{d\Delta}p(\Delta)$.

Paradoxically, the system converges to unstable solutions.  This
counterintuitive fact may be understood if we assume that the pair of
saddles $(p,\Delta')$ and $(1-p,\Delta'')$ are connected
heteroclinically \cite{okuda,hansel93}. All numerical results in
Ref. \cite{hansel93} support this assumption. Although the
heteroclinicity is considered structurally unstable, this does not seem
to apply to the particular class of systems under consideration.  In the
next section, it will be confirmed that $(p,\Delta')$ and
$(1-p,\Delta'')$ are in fact connected heteroclinically through an {\em
invariant subspaces}, and argued how this structure is maintained
stably.

\section{Structure of the heteroclinic loop} \label{sec:3}

We first note a particular symmetry of our phase model given by Eq.\ (1) 
which is expressed as 
\begin{equation} \label{identity}
 \left. \frac{d}{dt} \{\psi_i(t)-\psi_j(t) \} 
 \right|_{\psi_i(t)=\psi_j(t)}=0 \quad \mbox{for all $i,j$}. 
\end{equation}
The above equation tells when the phases of some oscillators are found
identical at some time, they will obey completely the same dynamics
thereafter or, equivalently, a point-cluster remains to be a
point-cluster forever.  This property will turn out crucial to the
formation of heteroclinic loops.  

We now argue how the phase model (\ref{pm}) can form heteroclinic loops
which is structurally stable.  A generalization to the vector model of
limit cycle oscillators will be given in the subsequent sections.
Existence of a heteroclinic loop connecting $(p,\Delta')$ and
$(1-p,\Delta'')$ is clear if the following properties are satisfied:
\begin{description}
 \item[(X)] $(p,\Delta')$ is a global attractor of 
	    $W^{\rm u}(1-p,\Delta'')$,
 \item[(Y)]  $(1-p,\Delta'')$ is a global attractor of
	    $W^{\rm u}(p,\Delta')$,
\end{description}
where $W^{\rm u}(p,\Delta)$ represents the unstable manifold of
$(p,\Delta)$.  We work with an $(N-1)$-dimensional phase space
throughout, by which the degree of freedom associated with a rigid
rotation of the entire system is ignored. For an aid to the
understanding of a little complicated situation, it would be appropriate
to display in advance a schematic picture of the heteroclinic loop under
consideration in Fig.\ \ref{fig:hetero}, where $\lambda'_i$,
$\lambda''_i$ and $\lambda'''_i$ $(i=1,2,3)$ are the eigenvalues of
$(p,\Delta')$, $(1-p,\Delta'')$ and $(p,\Delta''')$ respectively. The
definition of $E_{\rm X}$ (X=A,B,AB) will be given later.

Our argument will be based on the assumptions that there are three
two-cluster states in the range $p_{\rm min}<p<1-p_{\rm min}$ and that
the eigenvalues associated with these
solutions satisfy certain
stability properties. Specifically, the assumptions may be summarized as
follows:
\begin{figure}
\epsfxsize=8.5cm\epsfbox{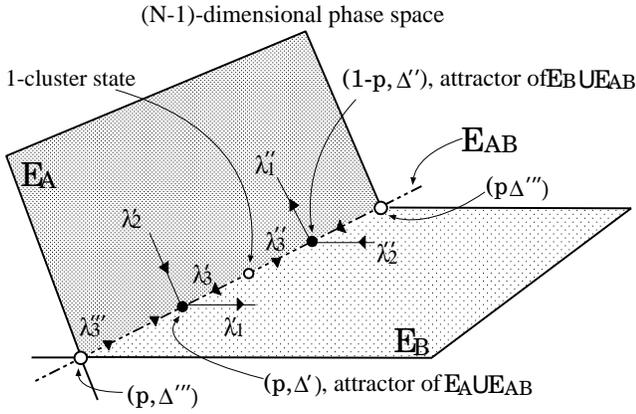}
  \caption{ Schematic representation of the structure of a heteroclinic
    loop.  $(p,\Delta')$ and $(1-p,\Delta'')$ are the attractors of the
    invariant subspace $E_{\rm A} \cup E_{\rm AB}$ and $E_{\rm B} \cup
    E_{\rm AB}$ respectively.  } \label{fig:hetero}
\end{figure}
\noindent
\begin{description}
 \item[(a)] $(p,\Delta')$, $(1-p,\Delta'')$ and $(p,\Delta''')$ exist, 
 \item[(b)]  $\lambda'_1>0$,
 \item[(c)]  $\lambda'_2<0$, $\lambda'_3<0$,
 \item[(d)]  $\lambda''_1>0$,
 \item[(e)]  $\lambda''_2<0$, $\lambda''_3<0$,
 \item[(f)]  $\lambda'''_3>0$,
 \item[(g)] $\lambda_1$ of all two-cluster states are positive,
 \item[(h)] $\Gamma'(x=0)>0$.
\end{description}
Here, we assumed that both $Np$ and $N(1-p)$ are larger than $1$
so that three independent eigenvalues $\lambda_i$ ($i=1,2,3$) exist.
Note that the whole assumptions are satisfied by Eq.\ (\ref{pm}) with
the coupling function given by Eq.\ (\ref{gamma-hansel}) as stated in
Sec. \ref{sec:2}.

Consider the tangent space around $(p,\Delta')$.  $(p,\Delta')$ has
$Np-1$ degenerate eigenvalues given by $\lambda'_1$. Thus, the
corresponding eigenvectors span a $(Np-1)$-dimensional unstable
eigenspace which is denoted by $E_{\rm B}$. Similarly, the eigenvectors
corresponding to $\lambda'_2$ span a $(N(1-p)-1)$-dimensional stable
eigenspace which is denoted by $E_{\rm A}$. An eigenvector corresponding
to $\lambda_3$ spans the $1$-dimensional stable eigenspace $E_{\rm
AB}$. In particular, the following statements hold.
\begin{description}
 \item[(b')]  $E_{\rm B}$ is the unstable subspace of $(p,\Delta')$,
\end{description}

$\lambda'_1$ corresponds to the fluctuations which occur in the
A-cluster (i.e. the phase-advanced cluster). Thus, the eigenspace
$E_{\rm B}$ is associated with the disintegration of the A-cluster,
while the B-cluster remains to be a point cluster. Similarly, the
B-cluster is disintegrated in the eigenspace $E_{\rm A}$, while the
A-cluster remains to be a point cluster there. In the space $E_{\rm
AB}$, in contrast, these clusters remain to be point-clusters while
their mutual distance changes.  Since a point-cluster must remain to be
a point-cluster at any time, as noted at the beginning of this section,
the space $E_{\rm B} \cup E_{\rm AB}$, on which the B-cluster is a
point-cluster, gives an {\em invariant subspace} of dimension
$Np$. Similarly, $E_{\rm A} \cup E_{\rm AB}$, on which the A-cluster is
a point-cluster, is an invariant subspace of dimension $N(1-p)$.  Note
that the unstable manifold $W^{\rm u}(p,\Delta')$ must coincide with
$E_{\rm B}$ in the vicinity of $(p,\Delta')$. This fact combined with
the obvious fact that $E_{\rm B}$ is included in the invariant subspace
$E_{\rm B} \cup E_{\rm AB}$ leads to the following statements which hold
globally:
\begin{description}
 \item[(x1)] $W^{\rm u}(p,\Delta')$ is included by the invariant subspace
	    $E_{\rm B} \cup E_{\rm AB}$.
\end{description}

Arguments parallel to the above can be developed around
$(1-p,\Delta'')$, i.e., the state where the B-cluster is phase-advanced
by $\Delta''$. From the assumed property (e), $E_{\rm B} \cup E_{\rm
AB}$ is the stable subspace of $(1-p,\Delta'')$, which can be restated
as follows:
\begin{description}
 \item[(e')] $(1-p,\Delta'')$ is an attractor of $E_{\rm B} \cup E_{\rm
	    AB}$.
\end{description}
Therefore, if  
\begin{description}
 \item[(x2)] $(1-p,\Delta'')$ is a {\em unique} attractor of $\rm E_B \cup
	    E_{AB}$,
\end{description}
then we may assume 
\begin{description}
 \item[(x3)] $(1-p,\Delta'')$ is a global attractor of $\rm E_B \cup
	    E_{AB}$.
\end{description}
(x1) and (x3) give the sufficient conditions for (X) to hold.  Similar
discussion can be developed to derive (Y) via the assumption
\begin{description}
 \item[(y2)] $(p,\Delta')$ is a {\em unique} attractor of $\rm E_A \cup
	    E_{AB}$.
\end{description}
Note that $(p,\Delta''')$ gives a source point lying between
$(p,\Delta')$ and $(1-p,\Delta'')$ in the phase space as displayed in
fig.\ \ref{fig:hetero}.

There still remains a problem about the validity of assumptions (x2) and
(y2). It is sufficient to consider (x2) only. If the type of the
attractors is limited to the two-cluster state, then it is obvious that
$(1-p,\Delta'')$ is a unique attractor of $E_{\rm B} \cup E_{\rm AB}$,
as can be confirmed by the property (g). How about the possibility for a
one-cluster state, namely, perfect synchrony, to become an attractor?
The eigenvalues of the one-cluster state are $N-1$-fold degenerate and
given by $K\Gamma'(x=0)$, which is positive by the assumption (h) so
that there is no stable manifold of one-cluster state. How about the
stability structure of $n(\geq 3)$-cluster states? They could be 
attractors of the invariant subspaces with the same reason as for the
two-cluster states even if they are unstable solutions. In the case of
three or more clusters, however, the resulting heteroclinicity would be
even more complicated. Numerical simulations in the previous section,
however, display the simple heteroclinicity between pairs of two-cluster
solutions, implying the validity of the assumption (x2) and (y2) in Eq.\
(1) with Eq.\ (2).

{\em Convergence} to the heteroclinic loop can be discussed similarly to
the case of the heteroclinic orbit in a two-dimensional phase space,
which was discussed in Ref. \cite{hansel93}. The result is that the
system which is initially close to a heteroclinic loop converges to it
provided
\begin{equation} \label{converge}
 \frac{\lambda_1' \lambda''_1}{\lambda'_2 \lambda''_2} < 1.
\end{equation}
If this condition is satisfied, the heteroclinic loop is attracting. In
numerical simulations, this convergence is established in a finite time
due to the round-off errors.

\begin{figure}
\centerline{\epsfxsize=7cm\epsfbox{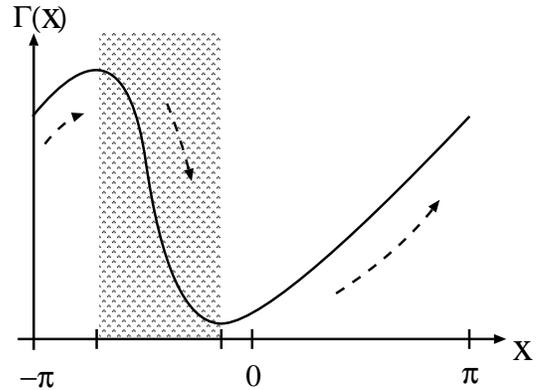}}
\caption{ Typical shape of the coupling function which admits a
  heteroclinic loop. This shape lead to conditions (a)-(h) except for
  $\lambda'_2, \lambda''_2<0$. The last property also hold if
  $\Gamma'(0)$ is not too large.} \label{fig:shape}
\end{figure}

Although we have assumed the conditions (a)-(h) so far, our discussion
is expected not to rely so heavily on the specific form of
$\Gamma(x)$. In fact, these conditions may be satisfied for a broader
class of the coupling function. For instance, they are satisfied if we
assume a simple shape of the coupling function such that $\Gamma(x)$
decreases in the range $-\pi < x < 0$, while it increases otherwise (see
Fig.\ \ref{fig:shape}). The reason is the following.  The corresponding
shape of $p(\Delta)$ turns out similar to the curve in fig.\
\ref{fig:p-delta-hansel}, so that we can define $p_{\rm min}$
similarly. For a given $p$ satisfying $p_{\rm min} < p < 1-p_{\rm min}$,
we obtain three states $(p, \Delta')$, $(p, -\Delta'') = (1-p,
\Delta'')$ and $(p,\Delta''')$. $\lambda_1$ of each $(p,\Delta>0)$ is
positive because $\Gamma'(0 \leq x \leq \pi) > 0$. We can verify
$\lambda'_3<0$, $\lambda''_3<0$ and $\lambda'''_3>0$ through the
property that $\lambda_3$ is proportional to $\frac{d}{d\Delta}
p(\Delta)$.  The one-cluster state turns out unstable since
$\Gamma'(0)>0$. Hence we have confirmed (a)-(h) except for $\lambda'_2,
\lambda''_2<0$. For the last properties to be satisfied, we need one
more assumption, that is, $\Gamma'(0)$ is not so large as to admit a
region of $p$ where both $\lambda'_2$ and $\lambda''_2$ are negative.
Such region is expressed by $p^* < p < 1-p^*$, where $p^*$ satisfies
$p_{\rm min} \leq p^* < 0.5 $. Then, via the assumptions (x2) and (y2),
we obtain sufficient conditions for the existence of a heteroclinic loop
between $(p, \Delta')$ and $(1-p, \Delta'')$ within the range $p^* < p <
1-p^*$.

In the previous works \cite{okuda,hansel93}, the shape of the
employed coupling functions satisfied this property.  Our model
discussed in the next section also fulfills this condition in the weak
coupling limit where the phase description is valid. Thus, we may regard
the coupling function with this property of the shape as a {\em typical}
class which admits heteroclinic loops.

The existence of the phase space structure yielding a heteroclinic loop
has thus been confirmed, which can be summarized as follows. For a given
coupling function $\Gamma(x)$, we can easily verify whether the
conditions (a)-(h) are satisfied. Among these conditions, (a)-(f)
constitute a necessary condition for the existence of a heteroclinic
loop between $(p,\Delta')$ and $(1-p, \Delta'')$, while (g) and (h)
support the assumptions (x2) and (y2). One may also consider the case
where the roles of $\lambda_1$ and $\lambda_2$ are reversed. The saddle
connections are stably formed through the invariant subspace which
exists for the symmetry of equations of motion, or
(\ref{identity}). Thus, we conclude that the heteroclinic loop is robust
under such small perturbations that maintain the homogeneity of the
population and the symmetry of the global coupling.

\section{Coupled limit cycle oscillators} \label{sec:4}
Our argument on the existence and structural stability of the
heteroclinic loop developed in the preceding section was based on the
phase model with some assumed properties of the phase-coupling function.
In this section, we discuss a specific coupled oscillator model in which
heteroclinic loops are formed. From the model, a phase-coupling function
of the desired properties for the existence of heteroclinic loops is
derived through the method of the phase reduction. To our knowledge, the
existence of the heteroclinic loop associated with slow switching has
never been reported for vector models of oscillators.

Consider a general system of coupled oscillators which are identical and
all-to-all coupled:
\begin{equation} \label{generalmodel}
\frac{d}{dt}{\hbox{$\boldmath X$}}_i(t) ={\bm F}({\bm X}_i)
                +\frac{K}{N} \sum_{j=1}^N {\bm G} ( {\bm X}_i, {\bm X}_j),
\end{equation}
Here $\bm{X}_i, \bm{F}$ and $\bm{G}$ are $m$-dimensional real vectors
and $K$ is a positive constant. Note that Eq.\ (14) satisfies the condition
\begin{equation} \label{general-ident}
 \left. \frac{d}{dt}\left\{{\bm X}_i(t) - {\bm X}_j(t) \right\}
 \right|_{{\bm X}_i(t) = {\bm X}_j(t)}   =  0 \qquad
 \mbox{for all \  $i,j$} ,
\end{equation}
which is similar to Eq.\ (12).  Suppose that the local dynamics is
two-dimensional, i.e. ${\bm X}=(x,y)$, and the specific forms of ${\bm
F}$ and ${\bm G}$ are given by
\begin{equation} \label{HRs}
{\bm F}({\bm X}_i) = \left( \begin{array}{l}
	          F_x \\
		  F_y
		 \end{array} \right)
        = \left( \begin{array}{l}
	          3x_i^{\ 2}-x_i^{\ 3}+y_i-\mu \\
		  1-5x_i^{\ 2}-y_i
		 \end{array} \right),
\end{equation}
\begin{equation} \label{coupling}
{\bm G}( {\bm X}_i, {\bm X}_j) = \left( \begin{array}{l}
	          G_x \\
		  G_y
		 \end{array} \right)
        = \left( \begin{array}{l}
	          x_j - x_i\\
		  0
		 \end{array} \right),
\end{equation}
\begin{figure}
\epsfxsize=8.5cm\epsfbox{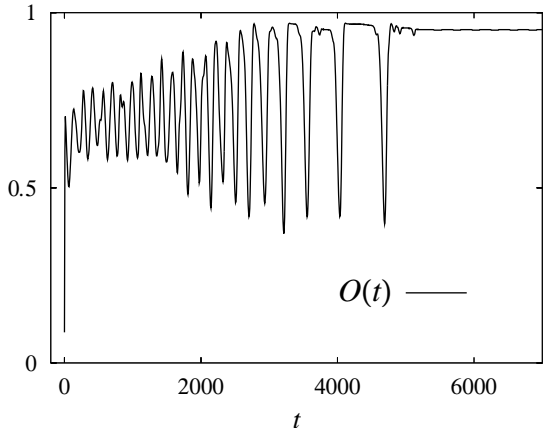}
  \caption{ Time series of the order parameter. The order parameter $O$
  is conveniently defined in the following way. Let $t_j$
  ($j=0,1,2,\cdots$) denote the time at which the representative point
  of the $j$-th oscillator crosses a given section $\Sigma$ in the
  $1$-oscillator phase space. The order parameter at time $t=t_N$ is
  defined as $ O(t=t_N) = \frac{1}{N} \left| \sum_{j=1}^N \exp \left[ i
  \frac{2 \pi (t_j-t_0)}{t_N-t_0} \right] \right|$ as a generalization
  of Eq.\ (3). Since the oscillators cross $\Sigma$ again and again, the
  order parameter at discrete times $t=t_{kN}$ ($k=1,2,\cdots$) may be
  defined similarly.  Note that $O(t)=1$ when the oscillators are
  synchronized perfectly and $O(t)=0$ when their phases are uniformly
  distributed.}  \label{fig:order-hr}
\end{figure}
\noindent
The corresponding equation $\dot{\bm X}={\bm F}$ is called the
Hindmarsh-Rose model \cite{hindmarsh} which was originally proposed for
a neural oscillator. Without coupling, i.e. $K=0$, each unit becomes
oscillatory if $-11.5 < \mu < 0.8$ \cite{hrwaves}. The coupling is
assumed to be diffusive, and in terms of neurophysiology, this
corresponds to the electrical synapse formed by gap junctions
\cite{MNM10}.

The parameter values are set to $K=0.1,N=100$ and $\mu=-1$. The
intrinsic frequency then becomes $\omega \simeq 1.0$. We choose random
initial conditions. Some numerical results obtained are summarized as
follows. The system converges after a long transient to a two-cluster
state which is periodic in time.  Figure \ref{fig:order-hr} displays a
time series of the order parameter.  The relative population of the
clusters generally depends on the initial condition.  Convergence to the
two-cluster state does not imply its stability, and is rather due to
numerical artifacts. Actually, when very small perturbations are given
to the oscillators independently, the clusters start to disintegrate,
implying their linear instability.  Such behavior is completely similar
to that of the phase model when heteroclinic loops exist. We now show
that the phase reduction of the above model produces a phase-coupling
function which admits, based on the argument of the preceding section,
heteroclinic loops.

\begin{figure}
\epsfxsize=8.5cm\epsfbox{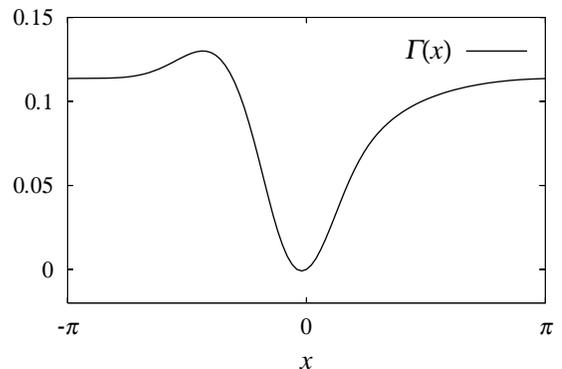}
  \caption{ Coupling function of the reduced model. The minimum of
  $\Gamma(x)$ appears at a negative $x$. Such a shape of $\Gamma$
  is {\em typical} when a heteroclinic loop exists.}
  \label{fig:gamma-hr}
\end{figure}

Coupled oscillators can be reduced to the phase model (\ref{pm}) when
the coupling is sufficiently weak \cite{kuramoto84}. There is a general
formula for the phase-coupling function, and for a given
dynamical-system model, this can be computed numerically. We did this
for Eqs.\ (\ref{generalmodel}), (\ref{HRs}) and (\ref{coupling}).  The
coupling function $\Gamma(x)$ obtained is displayed in 
Fig.\ \ref{fig:gamma-hr} which shows a typical shape admitting
heteroclinic loops (see Fig.\ \ref{fig:shape}). Two-cluster solutions
were sought and their stability analysis was done through Eqs.\
(\ref{p-delta})-(\ref{l3}). Then, we confirmed that the reduced model
satisfies the conditions (a)-(h) for the existence of a heteroclinic
loop and also the condition (\ref{converge}) for its stability.

\section{Structure of the heteroclinic loop in vector models} \label{sec:5}
The preceding arguments clarify the nature of the heteroclinic loop
in the framework of the phase model. In this section, we show that such
arguments can be generalized to the original model for coupled
limit-cycle oscillators in the vector form.

It would be appropriate to start reconsidering the model given by Eqs.
(\ref{generalmodel}), (\ref{HRs}) and (\ref{coupling}).  Under suitable
initial conditions, we obtain various two-cluster states. They
correspond to the solutions $(p,\Delta)$ with $\lambda_3<0$ in the
phase-coupling limit, and these two-cluster states will be denoted by
$p$-states. Also, the clusters corresponding to the phase-advanced and
retarded clusters will respectively be called the A-cluster and
B-cluster. For a given $p$-state, $mN$ Lyapunov eigenvalues can be
computed numerically, where $m=2$ for the model under
consideration. They can be classified into four groups $\Lambda_0$,
$\Lambda_1$s, $\Lambda_2$s and $\Lambda_3$s, and they will degenerate
respectively into $\lambda_i\ (i=0,1,2,3)$ in the phase-coupling limit.
Note that each of $\Lambda_1$s and $\Lambda_2$s is composed of $m$
eigenvalues, while $\Lambda_3$s is composed of $2m-1$ eigenvalues.
$\Lambda_1$s corresponds to the fluctuations within the A-cluster, and
each eigenvalue of this group is $Np-1$-fold degenerate. Similarly,
$\Lambda_2$s corresponds to the fluctuations within the B-cluster, and
each eigenvalue there is $N(1-p)-1$-fold degenerate. $\Lambda_3$s is
associated with the relative motion between the clusters.  $\Lambda_0$
is identical to $0$, resulting from the time-periodicity of the
solutions. The maximum value of $\Lambda_i$s

\begin{figure}
\epsfxsize=8cm\epsfbox{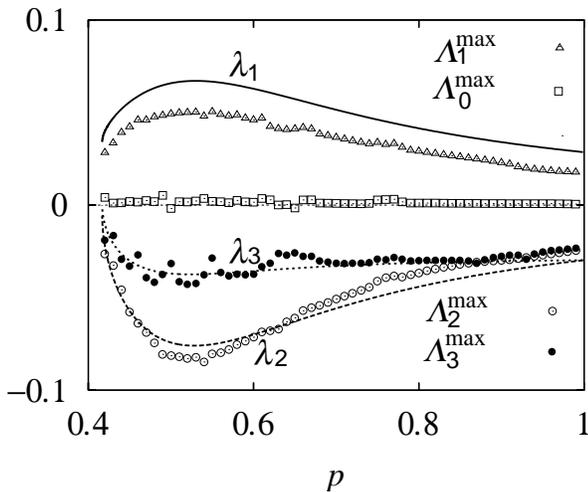}
  \caption{ Lyapunov spectrum plotted against $p$. From the figure,
  it can be read out that $p_{\rm min}
  \simeq 0.42$.  Solid lines show the eigenvalues obtained by 
  the phase reduction, with excellent agreement with the spectrum
  of the original system.}
  \label{fig:eigen-hr}
\end{figure}
\noindent 
($i=1,2,3$) is denoted by
$\Lambda_i^{\rm max}$, and their numerical values are displayed in Fig.\
\ref{fig:eigen-hr}.

The argument on the structure of a heteroclinic loop connecting the
$p$-state and $(1-p)$-state can be developed quite in parallel with that
in Sec. \ref{sec:3}. We have now to work with the $(mN-1)$-dimensional
phase space, whereby we employ a surface of section to remove the
irrelevant degrees of freedom corresponding to the steady rotation of
the whole population. The eigenspaces $E_{\rm A}$ and $E_{\rm B}$
associated with $\Lambda_1$s of $p$-state and $(1-p)$-state, are now
$m\{(N(1-p)-1\}$-dimensional and $m(Np-1)$-dimensional, respectively,
while the eigenspace $E_{\rm AB}$ associated with $\Lambda_3$s is
$(2m-1)$-dimensional.  Then the argument in Sec. \ref{sec:3} still holds
if we replace $\lambda_i$ with $\Lambda_i^{\rm max}$. Note that we
assume the existence of an unstable state corresponding to
$(p,\Delta''')$ which is hard to obtain numerically.

The eigenvalues $\Lambda_i^{\rm max}$ are the ones which should coincide
with $\lambda_i$ in the phase-oscillator limit. As the coupling becomes
stronger, the heteroclinic loop can persist as far as the existence- and
stability properties of the two point-clusters are unchanged.  Generally
speaking, the stronger coupling makes point clusters more stable.  In
Fig. 9, this effect is already sizable for $K=0.1$.  As $K$ becomes
$O(1)$, the two-cluster state gives way to a one-cluster state by which
the heteroclinic loop disappears. We expect in general that the
heteroclinic loop disappears when the coupling is so strong that the
phase description completely breaks down.

\section{Appearance of heteroclinic loops through delay-induced
 bifurcations} \label{sec:6} 

In globally coupled identical oscillators, one-cluster state is the
easiest state to appear. This can be illustrated by the following form
of coupling:
 \begin{equation} \label{coupling2}
{\bm G}( {\bm X}_i, {\bm X}_j) = {\bm X}_j(t) - {\bm X}_i(t). 
\end{equation}
Assuming Eqs. (\ref{generalmodel}), (\ref{HRs}) and (\ref{coupling2}),
we obtain a stable one-cluster state even if $K$ is very small.  We may
generally expect that the heteroclinic loop cannot exist when the
one-cluster state is stable. In the above model, it can be shown that
the time-delayed coupling causes instability of the one-cluster state,
which at the same time is accompanied by the appearance of the
heteroclinic loop. The corresponding bifurcation is the so-called
transcritical bifurcation.

Consider uniformly delayed coupling:
 \begin{equation} \label{coupling3}
{\bm G}( {\bm X}_i, {\bm X}_j) = {\bm X}_j(t-\tau) - {\bm X}_i(t) 
\end{equation}
where $\tau$ denotes the delay.  Note that the symmetry property
(\ref{identity}) still holds when the coupling involves a uniform delay
in the form of Eq. (\ref{coupling3}).  We will show some numerical
results obtained for the system given by Eqs.\ (14), (16) and (19) where
the parameter values are the same as in Sec. \ref{sec:4}.  Without
delay, the system under various initial conditions immediately converges
to a one-cluster state. As $\tau$ is increased, the one-cluster state
persists up to a critical value $\tau_c$ beyond which the cluster splits
into two and at the same time heteroclinic loops are formed. In this
case of the parameters, this critical value is about $0.18$.

This result can be understood by a phase reduction of our model which is
applicable when the coupling is weak.  The reduced model takes the form
\begin{equation} \label{pm-delay0}
 \frac{d}{dt}\psi_{i}(t) = \omega + \frac{K}{N}
	\sum^{N}_{j=1}\Gamma [ \psi_i(t)-\psi_j(t-\tau) ], 
\end{equation}
where $\omega\simeq 1.0$ at $\mu=-1$. Since the second term on the right-hand
side is much smaller than the first term by assumption, (\ref{pm-delay0}) is 
further reduced to the form 
\begin{figure}
\epsfxsize=8.5cm\epsfbox{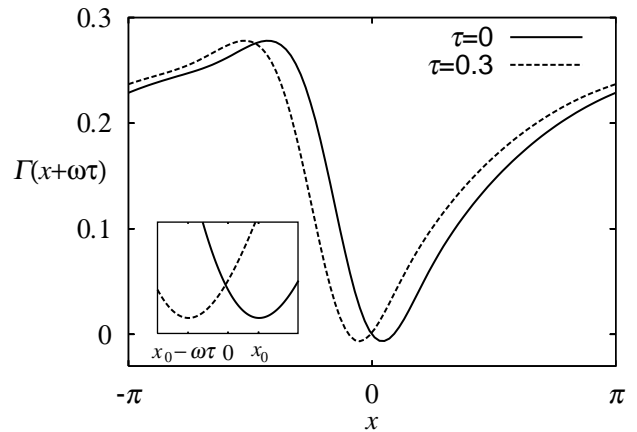}
  \caption{ Coupling functions. The solid line is obtained from
  (\ref{generalmodel}), (\ref{HRs}) and 
  (\ref{coupling3}) with $\tau=0$, while the dotted line is obtained
  just by a phase shift of solid line by $-\omega \tau$. 
  The effect of the delay is equivalent to a simple modification of the 
  coupling function in the weak-coupling limit.
  The modified coupling function is a typical shape admitting
  a heteroclinic loop provided the condition $\tau>
  x_0 / \omega$ is satisfied. } \label{fig:gamma-delay}
\end{figure}
\noindent
\begin{equation} \label{pm-delay}
 \frac{d}{dt}\psi_{i}(t) = \omega + \frac{K}{N}
	\sum^{N}_{j=1} \Gamma [ \psi_i(t)-\psi_j(t)+\omega \tau ].
\end{equation}
Thus, there is no explicit delay in coupling, while its effect has now
been converted to a phase shift of the coupling function by $\omega
\tau$.  The situation is illustrated if Fig.\ 10. The stability of a
one-cluster state depends entirely on the sign of $\Gamma'(\omega
\tau)$. Thus, the one-cluster state is stable for small $\tau$ which
admits $\Gamma'(\omega \tau)<0$.  As $\tau$ is increased, the
one-cluster state becomes less stable, and at $\tau = x_0 /\omega$ it
becomes unstable where $x_0$ is defined as the value of $x$ which
minimizes $\Gamma(x)$. Note that $\tau_c$ obtained numerically would
agree with $x_0 / \omega (\simeq 0.13)$ for sufficiently small $K$. For
$\tau> x_0 /\omega$, the coupling function assumes a typical shape under
which a heteroclinic loop exists. At $\tau=0.3$, for example, the
conditions (a)-(h) and (\ref{converge}) are satisfied in this reduced
model. This transition occurs through a transcritical bifurcation.

Let $x$ denote the mutual phase difference between the clusters. The
equation obeyed by $x$ can be derived similarly to the derivation of
Eq.\ (6), and takes the form
\begin{eqnarray} \label{pm-x-tau}
 \frac{d}{dt}x(t) = K\{ (2p-1)\Gamma(\omega \tau) 
  &+& (1-p)\Gamma(x+\omega\tau) \nonumber \\ 
 &-& p\Gamma(-x+\omega \tau) \}.
\end{eqnarray}
The above equation has a trivial solution $x=0$. For small $x$, the
right-hand side can be expanded in powers of $x$. We find that provided
$p \neq 0.5$ the right-hand side involves $x^2$ term as the lowest
nonlinearity.  This means that the trivial solution loses stability via
a transcritical bifurcation. This occurs at $\tau=x_0 / \omega$. We also
find that as $\tau$ is increased a saddle-node bifurcation occurs
slightly before the transcritical bifurcation. For $\tau>x_0 / \omega$,
the stable manifolds associated with the saddle-node and transcritical
bifurcations connect smoothly and forms a
\begin{figure}
\epsfxsize=8.5cm\epsfbox{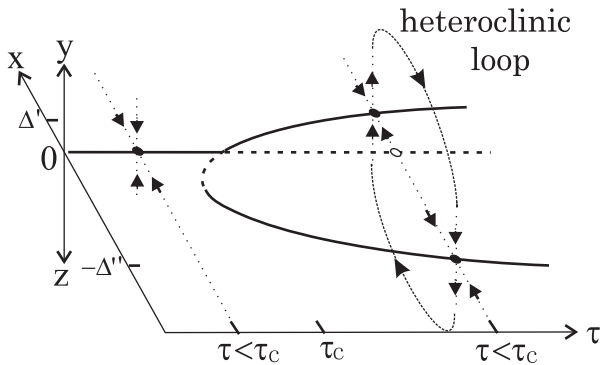}
  \caption{ Schematic representation of the bifurcation structure. 
 The trivial solution $x=0$ loses its stability at $\tau=\tau_c$ via a
 transcritical bifurcation. Two solid lines existing for $\tau>\tau_{\rm
 c}$ correspond to $(p,\Delta')$ and $(1-p,\Delta'')$ respectively, each 
 being unstable with respect to the $y$- or $z$-direction
 alternately. A heteroclinic loop is formed between these two solutions
 as explained in Sec. \ref{sec:3}} \label{fig:bunki}
\end{figure}
\noindent
heteroclinic loop.  Let $y$
(resp. $z$) denote a certain direction taken on $E_{\rm B}$
(resp. $E_{\rm A}$). How the bifurcation in question occurs is explained
schematically in Fig.\ \ref{fig:bunki}. 
The same structure of
bifurcation leading to the formation of heteroclinic loops holds for
some range of $p$ where $\lambda'_2$ and $\lambda''_2$ are both negative.
Note that at $p=0.5$ the term of $x^2$ vanishes so that a
pitchfork bifurcation occurs at $\tau=x_0 / \omega$ and no saddle-node
bifurcation occurs before. A heteroclinic loop is formed similarly to
the case of $p\ne 0.5$.

In coupled oscillators, the appearance of the heteroclinic loop may seem
to be pathological. The results in this section, however, imply that the
heteroclinic loop appears in a broad class of weakly coupled oscillators
including those studied so far provided the uniformly delayed coupling
is introduced. Assuming a simple coupling such as Eq. (18), we find that
when the assemblies are composed of relaxation oscillators, their phase
coupling function is often characterized by a curve which is sharply
decreasing in a small region while it is gradually increasing
otherwise. Thus, the heteroclinic loop is expected to arise in the
homogeneous assemblies of relaxation oscillators.

\section{Slow switching} \label{sec:7}
When the system involves heteroclinic loops, it exhibits a remarkable
dynamics when perturbed weakly. In the analysis of a model of the form
of Eq.\ (1), Hansel et al.\ \cite{hansel93} applied week noise
independently to the individual oscillators and observed the appearance
of a very long time scale depending on the noise intensity (see Fig.\
\ref{fig:slowswitch}). Since the time scale here is associated with the
alternation between two collective states (i.e. a pair of two-cluster
states), they called this characteristic behavior of the system {\em
slow switching}, and gave a successful explanation for it in terms of a
weakly perturbed heteroclinic loop.  Their explanation may be summarized
as follows.  If a heteroclinic loop is attracting, i.e. if Eq.\
(\ref{converge}) is satisfied, then the system approaches one saddle and
then to the other alternately. Without noise, the minimal distance from
each saddle should decrease exponentially in time.  With noise, however,
this distance will fluctuate but remain finite typically within the
order $\sigma$, the square-root of the variance of the noise.  In any
case, the system stays for the most time close to one saddle or the
other, so that the dynamics could be characterized dominantly by the
local properties around the saddles.  The time-interval $T$ for a stay
near a saddle may be estimated as
\begin{equation}
 T \sim - \frac{1}{\lambda_u} \ln \sigma,
\end{equation}
where $\lambda_u$ is the eigenvalue of the most unstable direction. 
Thus, the period of the switching is logarithmically dependent on the
noise intensity.

\begin{figure}
\epsfxsize=8.5cm\epsfbox{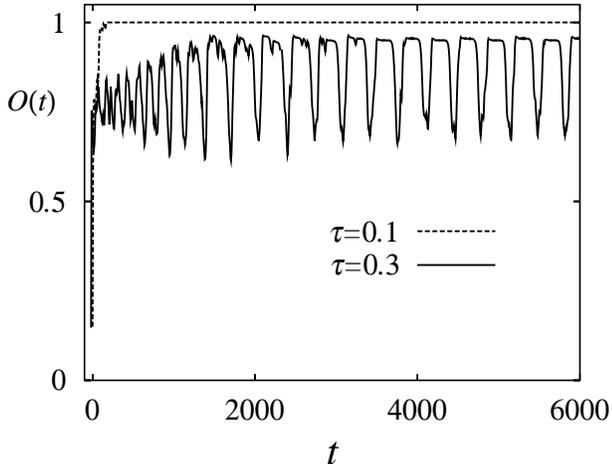}
  \caption{ Time series of the order parameter for a noisy system. 
  The solid line shows slow switching, where a new time scale of
  dynamics appears.}  \label{fig:hr-order-noise}
\end{figure}

\begin{figure}
\epsfxsize=8.3cm\epsfbox{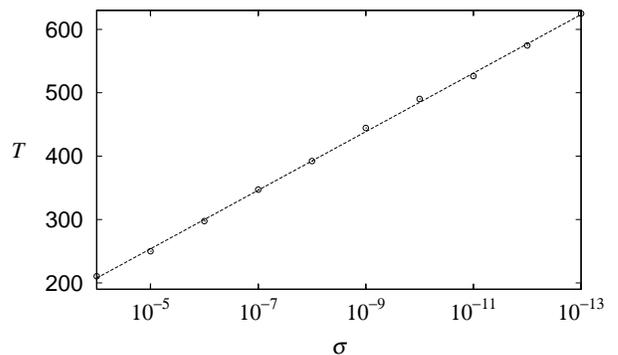}
  \caption{ Switching period versus noise intensity. The line is a 
  linear fitting of the data. }
  \label{fig:hr-period-delay}
\end{figure}

By including noise, Eq.\ (14) is generalized as
\begin{equation} \label{noisymodel}
\frac{d}{dt}{\bm X}_i(t) ={\bm F}({\bm X}_i)
 +K \sum_{j=1}^N {\bm G} ( {\bm X}_i, {\bm X}_j) 
 + \sigma\bm{\xi}_i (t),
\end{equation}
where $\bm{\xi}_i$ is Gaussian white noise with variance $1$ and the
parameter $\sigma$, assumed to be small below, indicates the intensity
of noise. Let us now consider specific forms for $\bm F$ and $\bm G$
given by Eq.\ (16) and Eq.\ (19), respectively, where the parameter
values are the same as before.  Some results obtained numerically are
the following. For $0 \leq \tau < \tau_{\rm c}$, the oscillators
localize in a small range of phase, which we call noisy one-cluster
state. Figure \ref{fig:hr-order-noise} displays a time trace of the
order parameter $O$, where $\tau=0.1$ and $\rho = 10^{-7}$. It is seen
that $O$ stays near $1$.  For $\tau>\tau_{\rm c}$, this highly
coherent cluster becomes unstable, and $O$ starts to oscillate.  After a
long transient, the system comes to exhibit slow switching between a
pair of two-cluster states.  For the most time the system stays close to
one of the noisy two-cluster state, which is followed by a short period
of cluster disintegration, then converging to another two-cluster state.
This is demonstrated in Fig.\ (14) for $\tau=0.3$.  It is clear that the
collective dynamics is then characterized by a new time scale
corresponding to this slow switching.  We define the period of switching
$T$ as the average time between the two successive local minima of $O$
sufficiently after the transient.  Logarithmic dependence of $T$ on the
noise intensity is clear from Fig.\ \ref{fig:hr-period-delay}. The
steepness of the $T$ versus $\ln \sigma$ curve after linear fitting is
estimated as $|T/\ln \sigma|\simeq 20$, which suggests that
$\Lambda_1^{\rm max}$ of this delayed coupling model is about
$0.05$. $\Lambda_1^{\rm max}$ can be easily estimated by Eq. (9) with
the phase-shifted coupling function displayed in Fig. 10. We obtain
$\lambda_1 \simeq 0.065$ at $p=0.5$, which is close to the
above estimation of $\Lambda_1^{\rm max} \simeq 0.05$. Note that
the amplitude effect makes $\Lambda_1^{\rm max}$ smaller than
$\lambda_1$ similarly to the case of Fig. 9.

Slow-switching is thus the fate of the system when the heteroclinic loop
is perturbed weakly. Besides external noise, a slight violation of the
symmetry property (\ref{general-ident}) is expected to cause the same
effect. Imagine a particular case where a heteroclinic loop is present
under the symmetry condition (\ref{general-ident}).  The system is now
perturbed slightly so that the condition (\ref{general-ident}) is
slightly violated.  Assume that a pair of saddles, between which a
heteroclinic loop is formed in the symmetric case, still exists.
Although a genuine heteroclinic loop could no longer exist in the
asymmetric system, the unstable manifold of one saddle will come close
to the other saddle.
If they are sufficiently close in the phase space, the situation is
quite similar to the case of applied weak noise, leading to
slow-switching.

\begin{figure}[bt]
\epsfxsize=8.5cm\epsfbox{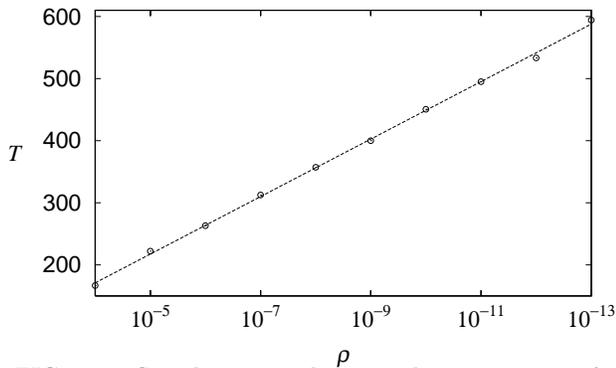}
  \caption{ Switching period versus the square-root of the
  variance of the parameter. The line is a linear fitting of the data.}
  \label{fig:hr-period-bunpu}
\end{figure}

As an example, let $\mu$ in Eq. (\ref{HRs}) be Gaussian-distributed with
variance $\rho^2$. Numerical simulations actually show slow switching
without noise.  Figure \ref{fig:hr-period-bunpu}, displaying the period
versus $\rho$, shows again a logarithmic law. The steepness of $|T/\ln
\rho|$ is estimated to be $20$, which implies that the gap $\epsilon$ is
now the order of $\rho$. Similar results are obtained when the other
parameters, e.g. the delay $\tau$, are randomly distributed. We can also
break the uniformity in the coupling, and consider a slightly random
oscillator network.

\section{Conclusions} \label{sec:8}

In the present paper, the physical relevance of the specific model
employed has little been discussed. We used the Hindmarsh-Rose
oscillator model which was originally proposed as a model for neuronal
oscillators. We have seen that the heteroclinic loop appears rather
easily in an assembly of relaxation oscillators, so that something
similar may be expected for homogeneous neuronal assemblies subject to a
constant external current. However, we did not intend modeling realistic
neuronal populations with the Hindmarsh-Rose model, but it was used just
for its simplicity in demonstrating the formation of the heteroclinic
loop in coupled oscillators especially when the coupling involves the
time-delay.  It could be understood
\noindent
from our whole argument that the same conclusion would be
valid for a much broader class of coupled oscillator models.  In
particular, any population model of coupled oscillators would be
capable to form the heteroclinic loop provided that it satisfies the
symmetry condition (\ref{general-ident}).
If we associate the switching phenomenon with biological rhythms, the
novel properties of this phenomenon would yield lots of suggestive
ideas. One of the intriguing subjects there is {\em rhythm splitting}
\cite{iglesia2000}. In connection with neural networks, a more realistic
model should be considered, and a study in this direction is now under
progress.

It may appear that the heteroclinic loop in question is something which
could not go beyond some mathematical curiosity because the symmetry
property (\ref{general-ident}) on which it crucially relies would be
more or less violated in real systems.  However, the associated
phenomenon of slow switching seems to be of much greater physical
relevance, because the strict symmetry need not be required there. Since
the noise, heterogeneity and delay are commonplace in macroscopic
systems, some indication of slow switching could well be detected in the
real world. In the case of mechanical oscillators, for instance, it
would not be difficult to obtain oscillators which are almost
identical. Global coupling might also be realized through electric
circuit \cite{kurt96}, vibrating board \cite{strogatz93}, surface motion
of water \cite{yoshikawa91} and so on. A certain class of surface
chemical reactions under oscillatory conditions may provide globally
coupled identical oscillators.

The characteristic frequency of slow switching will become shorter when
the symmetry breaking perturbations become stronger. At the same time,
the amplitude of the oscillating order parameter become smaller as the
perturbations become stronger. The switching phenomenon is expected to
vanish when the strength of the perturbation exceeds a critical value
after which the order parameter of the system becomes
stationary. Realistic examples of slow switching, if any, would
correspond to the case of strong perturbations. If slow switching
survives the strong perturbation and its frequency becomes comparable
with the intrinsic frequency of the oscillators, the dynamics would
become even more complicated due to the nonlinear coupling between these
modes of motion of comparable time-scales. Statistical mechanical
approach to this problem would be interesting. As far as we have
analyzed, however, slow switching vanishes well before its frequency
comes close to the intrinsic frequency. It would be also interesting to
find out coupled-oscillator models which have more robust structure of
slow switching with respect to the symmetry-breaking perturbations.

\section*{Acknowledgment}

The authors thank T. Mizuguchi and T. Chawanya for fruitful
discussions.

\end{document}